\documentclass[amsmath,amssymb,twocolumn,showpacs,reprint,aps,superscriptaddress,floatfix]{revtex4-1}
\usepackage{graphicx}
\usepackage{multirow}
\usepackage{dcolumn}
\usepackage{bm}

\newcounter{No}

\begin{document}
\title{Analysis of multinucleon transfer reactions involving spherical \\and statically deformed nuclei using a Langevin-type approach}
\author{V.~V. Saiko}
\affiliation{Flerov Laboratory of Nuclear Reactions, JINR, 141980 Dubna, Russia}
\affiliation{Dubna State University, 141982 Dubna, Russia}
\author{A.~V. Karpov}\email{e-mail: karpov@jinr.ru}
\affiliation{Flerov Laboratory of Nuclear Reactions, JINR, 141980 Dubna, Russia}
\affiliation{Dubna State University, 141982 Dubna, Russia}

\begin{abstract}
\begin{description}
    \item[Background] {An increasing interest in multinucleon transfer processes in low-energy deep inelastic (damped) collisions of heavy ions appeared in recent years is due, in part, to by the possibility of using them as a method of production of heavy neutron-enriched nuclei. Possible promising projectile-target combinations include nuclei deformed in the ground state (e.g., actinides). Mutual orientations of such the nuclei in the entrance channel of the reaction may significantly influence the reaction dynamics.}
    \item[Purpose] {A major aim of the work is to implement a possibility of modeling collisions of statically deformed heavy nuclei within a multidimensional dynamical model based on the Langevin equations. Another purpose of the paper is to study the influence of mutual orientation of statically deformed nuclei on their collision dynamics. Finally, the production yields of heavy transuranium nuclei in collisions of actinides are examined.}
    \item[Method] {The analysis has been performed within multidimensional dynamical model of nucleus-nucleus collisions based on the Langevin equations [Phys. Rev. C {\bf 96} 024618 (2017)]. In the present paper the model has been improved to describe collisions of statically deformed heavy nuclei with different mutual orientations.
        }
    \item[Results] {Available experimental data on multinucleon transfer reactions with statically deformed as well as spherical heavy nuclei $^{144}$Sm~+~$^{144}$Sm, $^{154}$Sm~+~$^{154}$Sm, $^{160}$Gd~+~$^{186}$W, and $^{208}$Pb~+~$^{208}$Pb/$^{238}$U have been analyzed within the developed model. A good agreement of the calculated quantities with the corresponding experimental data is reached. Special attention in the paper is paid to analysis of production possibility of the neutron-enriched isotopes of heavy and superheavy elements in multinucleon transfer processes in the $^{238}$U~+~$^{238}$U/$^{248}$Cm/$^{254}$Es collisions.
        }
    \item[Conclusions] {The mutual orientation of colliding statically deformed nuclei in the entrance channel strongly affects the energy, angular, mass, and charge characteristics of multinucleon transfer reaction products at near-barrier energies. These orientational effects disappear with increasing collision energy well above the Coulomb barrier. The exponential drop of the isotopic distributions of above-target products formed in the collisions of actinides with increasing atomic number does not allow one to synthesize new isotopes of superheavy nuclei with experimentally reachable cross sections. However, there is a possibility of production of a number of neutron-enriched isotopes of heavy actinides with the cross sections exceeding 1~${\rm \mu}$b.}
\end{description}
\end{abstract}
\maketitle


\section{Motivation}

Intensive nucleon transfer between interacting nuclei may occur in quasifission and deep inelastic (DI) collisions of heavy ions. This feature of DI collisions has already been used for synthesis of several light neutron-enriched nuclides shortly after the discovery of this type of nuclear reactions~\cite{Artukh73}. Nowadays a possibility of production of the neutron-enriched isotopes of heavy elements in these reactions is widely discussed~\cite{Dasso94,Broda06,ZagrebaevGreiner07_GdW,ZagrebaevGreiner08,Corradi09,Karpov17}. Recent experiments have confirmed applicability of this method~\cite{Kozulin12,Barrett15,Watanabe15,Kozulin17} in spite of the fact that a proper separation of the heavy nuclides from all multinucleon transfer (MNT) reaction products in a wide range of masses and kinetic energies still remains a challenging task.

Investigation of heavy neutron-rich nuclei in the vicinity of the $N=126$ neutron shell closure and study of their properties are of special interest, first of all, due to their influence on the problem of origination of chemical elements heavier than iron in the $r$-process of astrophysical nucleosynthesis.
The possibility of synthesis of neutron-rich nuclei with $N=126$ has been analyzed in our earlier paper~\cite{Karpov17}, where some of promising combinations of heavy nuclei were investigated within a multidimensional dynamical model based on Langevin equations. A reasonable agreement with available experimental data on angular, energy, mass, and charge distributions of MNT reaction products has been achieved. Cross sections for production of nuclei with $N=126$ have been predicted in a wide range of collision energies.

The systems studied in Ref.~\cite{Karpov17} involved heavy nuclei having spherical shapes in the ground state. However, collisions of heavy statically deformed nuclei are of particular interest as well. First, mutual orientation of the reaction partners influences the collision dynamics. There are some indications that the orientation of colliding nuclei at the contact point strongly affects the probability of formation of a compound nucleus in fusion reactions~\cite{Iwamoto96,Hinde96,Nishio00,Mitsuoka00}. This probability significantly decreases for nose-to-nose orientations, which leads to the decay of the nuclear system primarily in quasifission channels. Second, the collisions of deformed nuclei can be used for synthesis of new heavy or even superheavy neutron-enriched nuclei. For example, a possibility of production of neutron-rich nuclei in the vicinity of $N=126$ shell closure in the reaction of two statically deformed nuclei $^{186}$W + $^{160}$Gd has been predicted in Ref.~\cite{ZagrebaevGreiner07_GdW}. The orientational effects for this system of nuclei have been recently studied experimentally in Ref.~\cite{Kozulin17} and theoretically in Ref.~\cite{SaikoKarpov18}.

Actinides are a particular example of statically deformed nuclei. Obviously, formation of a compound nucleus is impossible in reactions between them. However, these reactions may lead to formation of neutron-enriched heavy and superheavy nuclei in DI collisions if the contact time of two nuclei is sufficient for transfer of a large number of nucleons. It may give one a possibility of production of several new neutron-enriched isotopes of transuranium elements with rather large cross sections. The yields of heavy nuclei up to mendelevium have been measured in the $^{238}$U~+~$^{238}$U/$^{248}$Cm reactions by radiochemical methods in Refs.~\cite{Schadel78,Schadel82}. Dynamics of $^{238}$U~+~$^{238}$U collisions has been studied experimentally in Ref.~\cite{Golabek10}. Extensive theoretical investigations of the collisions of actinides performed in the works of Zagrebaev and Greiner~\cite{ZagrebaevGreiner08,Zagrebaev11} have shown that strong shell effects in the potential energy of the nuclear system favor the formation of the first fragment in the region of doubly magic lead and the second one in the region of superheavy elements. Moreover, the potential energy structure favors the formation of neutron-enriched nuclei unreachable in fusion reactions of stable nuclei. Recently, this problem has been also investigated in Refs.~\cite{Zhao13,Zhao16} within the improved quantum molecular dynamics model as well as in Ref.~\cite{Feng_UU09} within the dinuclear system model.

Consideration of the orientational degrees of freedom in dynamical calculations within any model using the concept of nuclear shape is a quite complicated problem and has no straightforward solution yet. Reasonable assumptions should be made in order to avoid some major difficulties. The main problem here is the calculation of the potential energy corresponding to sequential shapes of the nuclear system, which are unknown in absence of the axial symmetry. This concerns in particular the mechanism of restoration of the axial symmetry of two arbitrary oriented nuclei at the contact point.

Thus, the first aim of this work is to advance the model of Ref.~\cite{Karpov17} in order to simulate collisions of heavy statically deformed nuclei. The second purpose of the paper is to study the impact of mutual orientation of statically deformed nuclei on their collision dynamics. Finally, the production yields of heavy transuranium nuclei in collisions of actinides are examined.

\section{\label{sec:model}Model}

The model has been described in detail in Ref.~\cite{Karpov17}. Here we provide only its basics. The model has eight degrees of freedom describing a system of two colliding nuclei. Four of them originate from the well-known Two-Center Shell Model (TCSM) parametrization~\cite{Maruhn72} and define the nuclear shape: distance~$r$ between geometrical centers of two nuclei, two independent ellipsoidal surface deformations~$\delta_{1,2}$, and mass asymmetry~$\eta_A$. In addition, the charge asymmetry~$\eta_Z$ is included in the model that allows us to describe independent proton and neutron transfer and production of nuclei with different atomic and mass numbers. Two angles~$\varphi_{1,2}$ of rotation of the projectile and target nuclei and the angle~$\theta$ between the internuclear axis and the beam direction are included in the model as well.

A set of eight coupled Langevin equations is solved numerically:
\begin{eqnarray}\label{eq:Langevin}
&&\dot q_i = \sum_j \mu_{ij} p_j, \nonumber \\
&&\dot p_i = T\left(\frac{\partial S}{\partial q_i}\right)_{E_{\rm tot}} -\sum_{j,k} \gamma_{ij} \mu_{jk} p_{k} + \sum_j \theta_{ij} \xi_j(t),\label{eq:Langevin}
\end{eqnarray}
where $q_i=\{r,\delta_1,\delta_2,\eta_A,\eta_Z,\theta,\varphi_1,\varphi_2\}$ and $p=\{p_r,p_{\delta_1},p_{\delta_2},p_{\eta_A},p_{\eta_Z},L,l_1,l_2\}$ are the collective degrees of freedom and their conjugate momenta, respectively. $S=2\sqrt{aE^*}$ is the entropy of excited system, where $a$ is the level density parameter and $E^*=E_{\rm tot}-V-E_{\rm kin}$ is the excitation energy. Here $E_{\rm tot}$ is the total energy of the system, $E_{\rm kin}$ is the kinetic energy stored in all collective degrees of freedom, and $V$ is the potential energy. $\mu_{ij}=[m_{ij}]^{-1}$  is the inverse inertia tensor, $\gamma_{ij}$ is the friction tensor, $\theta_{ij}$ are the amplitudes of the random forces determined from the Einstein equation $\theta_{ik}\theta_{kj}=\gamma_{ij}T$, $\xi_i$ are the normalized random variables with Gaussian distribution $\langle\xi_i(t)\rangle = 0$, $\langle \xi_i(t),\xi_{j}(t') \rangle = 2\delta_{ij}\delta(t-t')$, and $T=\sqrt{E^*/a}$ is the nuclear temperature. The terms in the Eq.~(\ref{eq:Langevin}) represent respectively the driving, friction, and random forces.

The potential energy, friction, and inertia coefficients are the main values that govern the evolution of the nuclear system. We calculate them on a grid before starting the dynamical calculations. The multidimensional potential energy takes into account transition from the diabatic regime of nuclear motion to the adiabatic one as a relaxation process with the relaxation time $\tau_{\rm DA}$:
\begin{eqnarray}\label{eq:Vpot}
V(r,\delta_1,\delta_2,\eta_A,\eta_Z;t) = &&V_{\rm diab}\exp\left(-\frac{t}{\displaystyle \tau_{\rm DA}}\right) +\nonumber\\
+ &&V_{\rm adiab}\left[1-\exp\left(-\frac{t}{\displaystyle \tau_{\rm DA}}\right)\right].
\end{eqnarray}
The diabatic potential is calculated within the double-folding method with Migdal nucleon-nucleon forces~\cite{Migdal83}. The extended macro-microscopic approach~\cite{ZagrebaevKarpov07,NRV} is used to calculate the adiabatic potential:
\begin{eqnarray}\label{eq:Vadiab}
V_{\rm adiab}(r,\delta_1,\delta_2,&&\eta_A,\eta_Z) = \nonumber\\
&&V_{\rm mac}(r,\delta_1,\delta_2,\eta_A,\eta_Z) + \delta E(r,\delta_1,\delta_2,\eta_A,\eta_Z), \nonumber
\end{eqnarray}
where macroscopic term $V_{\rm mac}$ is calculated within the finite-range liquid-drop model~\cite{Moller16}, and $\delta E$ is the shell correction calculated based on the Strutinsky method~\cite{Strutinsky67, Strutinsky68}.

We calculate the inertia coefficients for the $r,\delta_1,\delta_2,\eta_A$ degrees of freedom according to the Werner-Wheeler approach for incompressible irrotational flow~\cite{Davies76}. The corresponding friction coefficients are calculated within the ``wall+window'' mechanism of one-body dissipation~\cite{Sierk80}. The inertia and friction coefficients for the charge asymmetry are calculated in the same manner as in Ref.~\cite{KarpovEPJA02}.

Modeling nucleon transfer and dissipation of the energy of relative motion before the contact of nuclei in the entrance channel of the reaction requires non-zero friction and inertia coefficients for the corresponding degrees of freedom. This is achieved by introducing additional phenomenological form-factors with adjustable parameters.

Solution of Eq.~(\ref{eq:Langevin}) starts from the initial distance between nuclei $\approx$50 fm. A projectile with a given impact parameter $b$ and a certain center-of-mass energy~$E_{\rm c.m.}$ approaches a target nucleus. The nucleon transfer and energy dissipation processes start to occur slightly before the nuclei come into contact. Then the dinuclear system forms and decays into two excited reaction fragments. The calculations are terminated when the products are formed and separated again by the initial distance. The obtained solution is a trajectory in the multidimensional space of the collective degrees of freedom that carry complete information about a single collision.

A large number of trajectories for different impact parameters are simulated. Then the statistical model~\cite{Karpov17,NRV,KarpovNRV16,KarpovNRV17} is used to obtain final reaction products from the primary excited ones. Usually the Monte-Carlo method of simulation is used in calculations within the statistical model, but for highly-fissile reaction products the method of nested integrals is applied under the assumption that final products are formed in neutron evaporation channels (up to four neutrons). The GEF code is employed for simulation of sequential fission (SeqF) fragment distribution~\cite{GEF}. Finally, the cross sections are calculated as:
\begin{eqnarray}
  \label{eq:cross-section}
  \frac{d^4\sigma}{dZdAdEd\Omega}&&(Z,A,E,\theta)= \\
  &&\int\limits_0^{b_{\rm max}} \frac{\Delta N(b,Z,A,E,\theta)}{N_{\rm tot}(b)}\frac{b\,db}{\Delta Z \Delta A \Delta E \sin{\theta} \Delta \theta},\nonumber
\end{eqnarray}
where $\Delta N$ is a number of trajectories in a given bin and $N_{\rm tot}$ is the total number of simulated trajectories for each impact parameter. Integration of Eq.~(\ref{eq:cross-section}) allows one to obtain different distributions of reaction products.

In order to consider the collisions of statically deformed nuclei one should take into account that the reaction dynamics is strongly dependent on the mutual orientations of the nuclei. This happens, first of all, because of the change of the potential energy. The orientation effects in the potential energy can be easily considered for separated ``frozen'' nuclei. In this case the potential energy is calculated using the double-folding procedure~\cite{Migdal83}. After the contact, the potential energy depends on the interaction time. In particular, the axial symmetry should be restored if the interaction time is long enough. Another difficulty in calculation of the potential energy of a strongly interacting nuclear system with broken axial symmetry is connected with appearance of dynamical deformations. They should develop predominately along the axis connecting geometrical centers of the interacting nuclei. This destroys axial symmetry of each of the reaction partners that complicates the task even more.

Thus, the sequence of shapes passed by arbitrary oriented colliding nuclei in the vicinity of the contact point is unknown and the calculation of the corresponding potential energy for these shapes with broken axial symmetry is rather complicated and yet-unsolved problem. It is difficult or even impossible to solve this problem based on the implying any parametrization of nuclear shape without substantial increase of a number of degrees of freedom. Finally, one may conclude that a certain approximation should be employed.

Detailed studies within the time-dependent microscopic approaches may shed more light on the problem of evolution of nuclear shapes for arbitrary oriented interacting nuclei.
For example, according to time-dependent calculations for the $^{48}$Ca~+~$^{249}$Bk collisions with berkelium orientation of 45$^{\circ}$ it takes the dinuclear system about 3$\cdot10^{-21}$ s to restore its axial symmetry~\cite{Oberacker14,Umar16}. We suppose that this time should be close to the period of quadrupole oscillations of an isolated nucleus. The corresponding value for the $^{249}$Bk nucleus is approximately 6$\cdot10^{-21}$ s, which is in line with this assumption.

\begin{figure}[t]
\centering
  \includegraphics[width=1\linewidth]{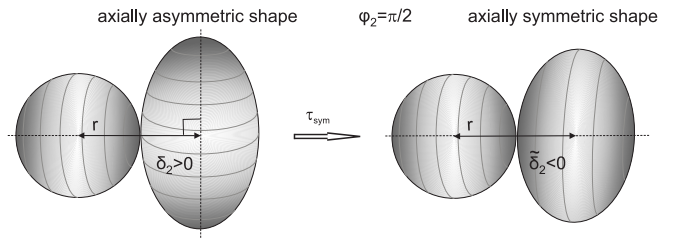}
  \caption{Example of evolution of axially asymmetric nuclear system to axially symmetric one.}
  \label{img:shapes}
\end{figure}

It is rather natural to use the exponential factor in order to simulate the process of restoration of axial symmetry in the potential energy. The potential energy is thus calculated as
\begin{equation}
V = V_{\rm asym} \exp{\left(-\frac{t}{\tau_{\rm sym}}\right)} +
V_{\rm sym} \left[1 - \exp{\left(-\frac{t}{\tau_{\rm sym}}\right)} \right],
\end{equation}
where $V_{\rm sym}$ and $V_{\rm asym}$ are the potentials (\ref{eq:Vpot}) calculated for axially symmetric and axially asymmetric configurations, respectively, and $\tau_{\rm sym}$ is the relaxation time. We have found that the value of this relaxation time mainly influences the maximal value of the kinetic energy loss in deep inelastic collisions of heavy ions. Larger values of $\tau_{\rm sym}$ lead to smaller maximal kinetic energy losses, because the system longer experiences the potential energy for oriented nuclei having larger barrier than the one for axially symmetric configuration. The value used in the present calculation was chosen as described above to be $\tau_{\rm sym}= 10^{-21}$~s.

We assume that the relaxation process of an arbitrary oriented nuclear system to an axially symmetric one starts when the colliding nuclei approach the distance of strong interaction ($\sim2$ fm between nuclear surfaces). The relaxation process cannot result in an increase of the potential energy of the system. The potential energy of two heavy nuclei at the vicinity of the contact point is mainly determined by the distance between the nuclear surfaces. Thus, we assume that arbitrary oriented nuclei keep the distance between their surfaces (as well as the relative distance $r$) while the relaxation process proceeds in the perpendicular direction (see Fig.~\ref{img:shapes}). This allows us to relate the ellipsoidal deformations $\delta_i$ of initially oriented pair of colliding nuclei with the ellipsoidal deformations $\tilde{\delta_i}$ of final axially symmetric shapes:
\begin{equation}
 \tilde{\delta_i} = \left(1+\delta_i\right)\left[\delta_i(2+\delta_i)\sin^2\varphi_i+1\right]^{-\frac{3}{4}} -1.
\end{equation}

It is often assumed that initial mutual orientations of colliding nuclei do not change before the nuclei come into the contact. This assumption is also used in the present model, which allows us to simplify the consideration of collision dynamics for separated nuclei substantially. However, time evolution of the angles $\varphi_i$ cannot be ignored completely. In the vicinity of contact point, sliding friction between nuclear surfaces leads to the so-called dissipation of the angular momentum of relative motion until the sticking condition is reached. Part of the relative angular momentum is thus transferred into the angular momenta of the reaction partners that results in a time evolution of the angles $\varphi_i$, which should be, therefore, kept in the system of the Langevin equations (\ref{eq:Langevin}). Such consideration allows us to simulate nonzero momenta of the reaction products.

In order to simplify the calculations, we have considered only limit initial orientations of two deformed colliding nuclei: the so-called tip-to-tip ($\varphi^0_1=\varphi^0_2=0$), side-to-side ($\varphi^0_1=\varphi^0_2=\pi/2$), tip-to-side ($\varphi^0_1=0,\varphi^0_2=\pi/2$), and side-to-tip ($\varphi^0_1=\pi/2,\varphi^0_2=0$) collisions. If the orientation of the $i$th nucleus is $\varphi^0_i=\pi/2$, then the corresponding deformation of the axially symmetric shape has a simple form:
\begin{equation}
 \tilde{\delta_i} = \left(1+\delta_i\right)^{-\frac{1}{2}} -1.
\end{equation}

Finally, the cross sections are averaged over initial mutual orientations as:
\begin{eqnarray}
  \label{eq:aver}
  &&\langle \sigma(Z,A,E,\theta)\rangle = \nonumber\\
  &&\int\limits_0^{\frac{\pi}{2}}\int\limits_0^{\frac{\pi}{2}} \sigma(Z,A,E,\theta,\varphi_1^0,\varphi_2^0) \sin\varphi_1^0 \sin\varphi_2^0 d\varphi_1^0 d\varphi_2^0,
\end{eqnarray}
where the cross section $\sigma(Z,A,E,\theta,\varphi_1^0,\varphi_2^0)$ is calculated at four limit orientations of colliding nuclei $\varphi_i^0=0, \pi/2$ and it is assumed to be linearly dependent on the angles $\varphi_1^0$ and $\varphi_2^0$.

\section{Analysis of multinucleon transfer reactions}

In our previous work~\cite{Karpov17}, the developed model was successfully applied to analysis of reaction dynamics and production of heavy neutron-enriched nuclei in MNT reactions with heavy spherical nuclei $^{136}$Xe~+~$^{198}$Pt/$^{208}$Pb/$^{209}$Bi. The present work is devoted to comparison of DI collisions involving deformed and spherical nuclei. All the calculations were done within a set of the model parameters fixed in Ref.~\cite{Karpov17}.

The discussion of the studied reactions is ordered by the mass of the composite systems. We demonstrate how accurately the developed model is able to describe various experimental measurable characteristics. The influence of mutual orientation of statically deformed nuclei of these characteristics is also discussed. Finally, we analyze the possibility of production of neutron-enriched transuranium nuclides in low-energy collisions of actinides.

\begin{figure}[t]
\centering
  \includegraphics[width=1\linewidth]{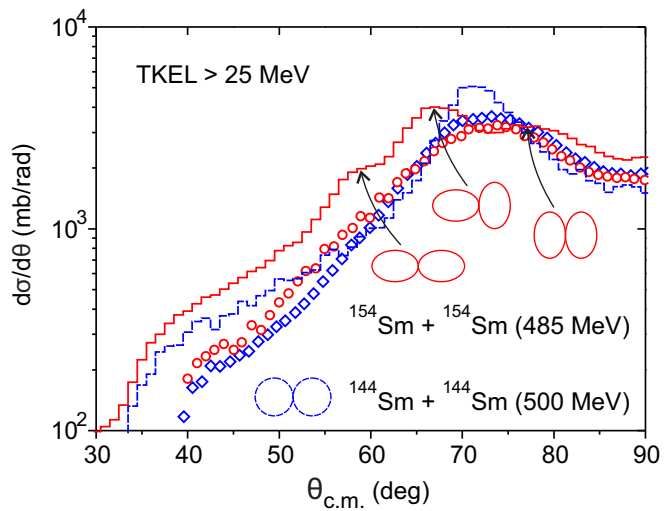}
  \caption{\small The angular distributions of reaction products with TKEL$\geq$25 MeV obtained in the $^{144}$Sm~+~$^{144}$Sm and $^{154}$Sm~+~$^{154}$Sm reactions at energies $E_{\rm c.m.}=500$ and 485 MeV, respectively. Histograms are calculation results and symbols show the experimental data~\cite{Hildenbrand83}. The dashed histogram and diamonds correspond to the $^{144}$Sm~+~$^{144}$Sm reaction. The solid histogram and circles are for the $^{154}$Sm~+~$^{154}$Sm reaction.}
  \label{img:SmAngles}
\end{figure}

\subsection{$^{144}$Sm~+~$^{144}$Sm and $^{154}$Sm~+~$^{154}$Sm systems}

\begin{figure}[b]
\centering
  \includegraphics[width=0.9\linewidth]{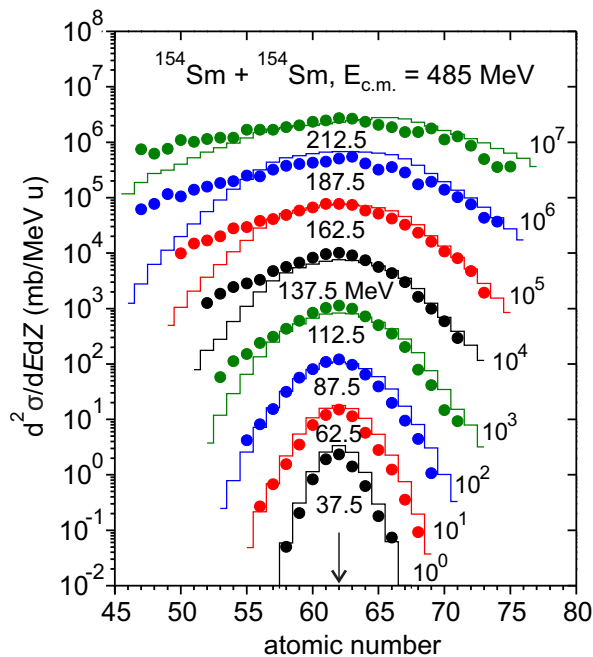}
  \caption{\small The charge distributions of final products for different values of TKEL obtained in the $^{154}$Sm~+~$^{154}$Sm reaction at $E_{\rm c.m.}=485$~MeV. The energy bins are 25 MeV wide. The midpoints of each bin are indicated. Both experimental and calculated data were multiplied by factors indicated on the side.}
  \label{img:SmCharges}
\end{figure}

The first combination consists of spherical magic nuclei ($\beta_2(^{144}$Sm$)=0$), while the second one involves well-deformed nuclei ($\beta_2(^{154}$Sm$)=0.27$~\cite{Moller16}). These two reactions are of particular interest because they allow us to study the influence of mutual orientations of colliding nuclei in absence of SeqF of highly excited reaction products. The differential cross sections of fragments obtained in the both reactions were measured under the same experimental conditions. The detectors covered the following angular ranges: $15^\circ \leq \theta_{\rm lab.} \leq 35^\circ$ and $27^\circ \leq \theta_{\rm lab.} \leq 47^\circ$. The comparable collision energies $E_{\rm c.m.}=500$ and 485 MeV were chosen respectively for the $^{144}$Sm~+~$^{144}$Sm and $^{154}$Sm~+~$^{154}$Sm reactions~\cite{WuHildenbrand81,Hildenbrand83}. The experimental charge resolution of 1.6 units (FWHM) was taken into account in the calculations.

\begin{figure*}[t]
\centering
  \includegraphics[width=1\linewidth]{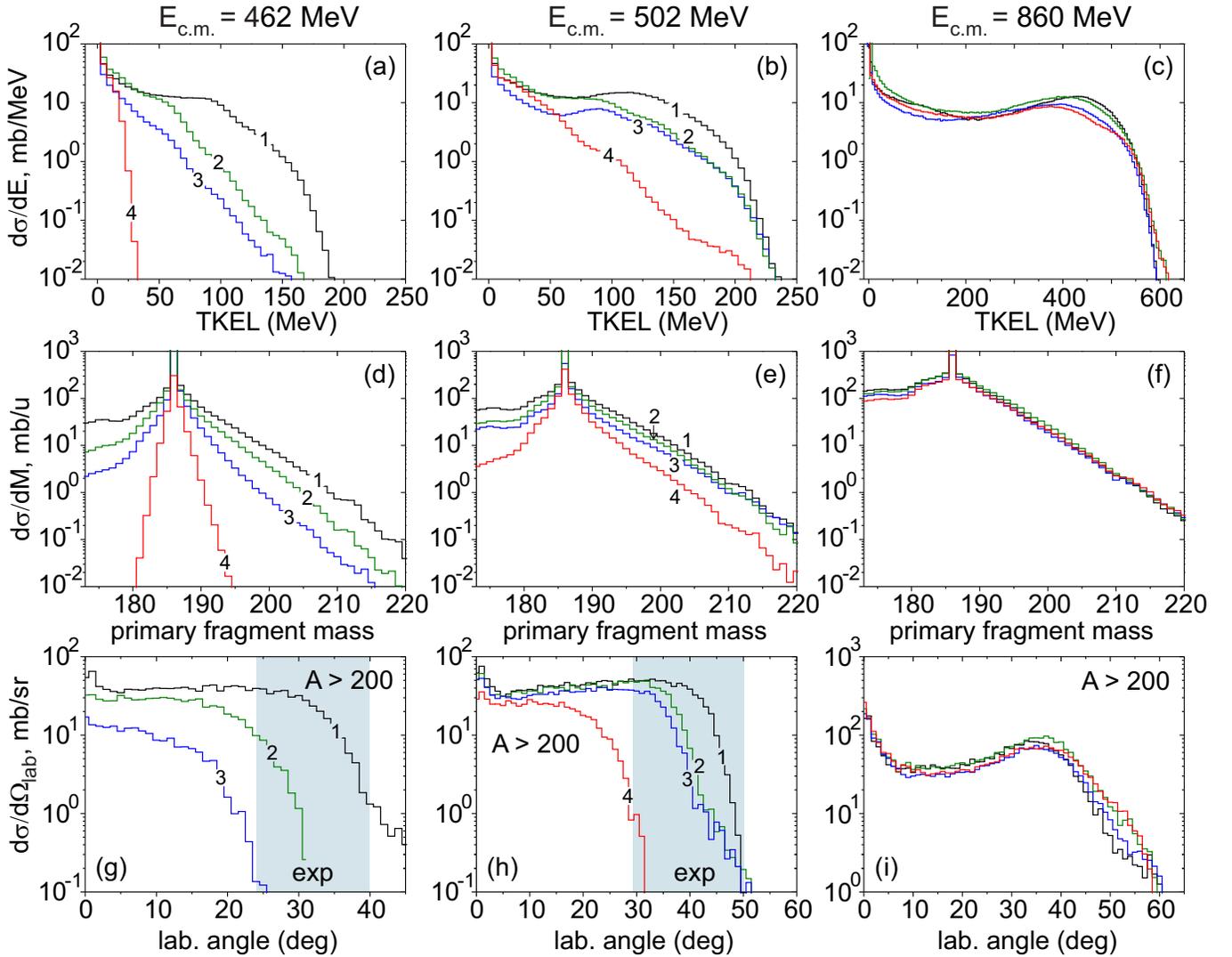}
  \caption{\small TKEL, mass, and angular distributions of primary products of the $^{160}$Gd~+~$^{186}$W reaction calculated at three energies $E_{\rm c.m.}=462$, 502, and 860 MeV. Histograms show results for tip-to-tip (1), tip-to-side (2), side-to-tip (3), and side-to-side (4) collisions. Shaded areas in the angular distributions show the angular ranges covered in the experiment for detecting TLFs~\cite{Kozulin17}.}
  \label{img:GdWorientations}
\end{figure*}

The angular distributions of final products of the Sm~+~Sm reactions with total kinetic energy losses (TKEL) larger than 25 MeV are shown in Fig.~\ref{img:SmAngles}. It is clear that the model provides a good description for the shape and the position of the maximum of the distribution for the $^{144}$Sm~+~$^{144}$Sm reaction involving spherical nuclei in their ground states. The angular distribution of fragments obtained in the reaction between deformed nuclei $^{154}$Sm~+~$^{154}$Sm also agrees reasonably with the data having, however, the width somewhat exceeding the experimental one. The shape of the distribution insignificantly suffers from the method of taking into account the orientational effects. Different orientations lead to different angles of grazing collisions. A more compact side-to-side configuration has larger grazing angle. The tip-to-side and side-to-tip orientations for this symmetric system lead to the same grazing angle in the center-of-mass frame. The model considers only four different mutual orientations of colliding nuclei. That is why the calculated curve has three maxima corresponding to the tip-to-tip, tip-to-side and side-to-side collisions. Better results for the angular distribution for the $^{154}$Sm~+~$^{154}$Sm reaction can be achieved by taking into account additional mutual orientations of nuclei in the averaging procedure or by a complete consideration of evolution of the orientational degrees of freedom in dynamical calculations.

Another important feature of the reaction dynamics is an increase of nucleon transfer with increased TKEL. It can be seen from Fig.~\ref{img:SmCharges} where the charge distributions of the $^{154}$Sm~+~$^{154}$Sm reaction products are plotted for different TKEL values. The experimental data and results of the calculations were both scaled for convenient visualization. The calculation results have gaussian form and fit well the experimental data. An underestimation for light products with $Z<62$ is due to the experimental background, which was not completely eliminated during the data analysis~\cite{Hildenbrand83}. The calculations predict more symmetrical charge distributions.

\subsection{$^{160}$Gd~+~$^{186}$W system}

\begin{figure*}[t]
\centering
  \includegraphics[width=0.9\linewidth]{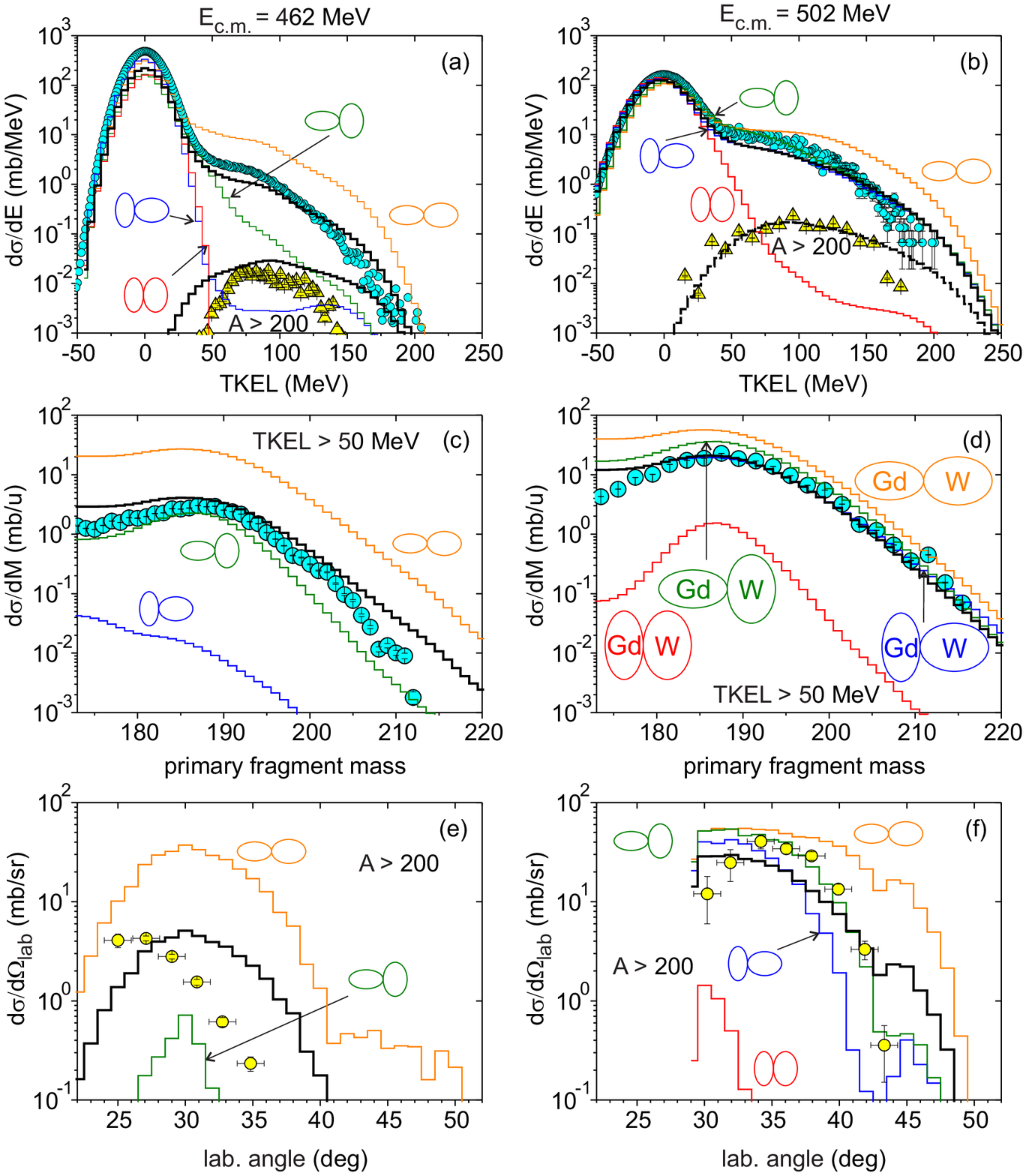}
  \caption{\small TKEL, mass, and angular distributions of primary products obtained in the $^{160}$Gd~+~$^{186}$W reaction at two energies $E_{\rm c.m.}=462$ and 502 MeV. Symbols are the experimental data~\cite{Kozulin17}, the thin lines are the calculations for tip-to-tip, side-to-tip, tip-to-side, and side-to-side collisions; the thick lines show the total cross sections averaged over the mutual orientations of the projectile and target nuclei. The angular distributions as well as the dashed histograms and triangles for the energy distributions correspond to TLFs with $A>200$.}
  \label{img:GdWresults}
\end{figure*}

This approach was applied to the $^{160}$Gd~+~$^{186}$W collisions, where both reaction partners have prolate deformations in the ground state $\beta_2(^{160}$Gd$)=0.28$ and $\beta_2(^{186}$W$)=0.22$~\cite{Moller16}. The experimental data for this reaction have been recently obtained at two near-barrier collision energies $E_{\rm c.m.}=462$ and 502 MeV~\cite{Kozulin17}. It allows one to study the influence of mutual orientations on the primary TKEL, mass, and angular distributions of reaction products calculated for the above mentioned energies (see Fig.~\ref{img:GdWorientations}). The angular distributions were calculated for target-like fragments (TLFs) with $A\leq200$.

The experimental collision energies are in the same energy range as the calculated Coulomb barriers for different orientations of this system, which are $V_C=412$~MeV for the tip-to-tip, $V_C=447$~MeV for the tip-to-side, $V_C=454$~MeV for the side-to-tip, and $V_C=492$~MeV for the side-to-side collisions. The strongest influence of the nuclear orientations on the collision dynamics is seen for the lowest energy ($E_{\rm c.m.}=462$ MeV) corresponding to the barrier for the tip-to-side and side-to-tip configurations. Side-to-side collisions are the sub-barrier ones at this energy. This strongly suppresses the nucleon transfer as well as the kinetic energy dissipation. The grazing angles and, thus, angular distributions are also significantly affected by the mutual orientations of the colliding nuclei.

The intermediate energy ($E_{\rm c.m.}=502$ MeV) is just above the barrier for the side-to-side collisions. The distributions of the reaction products become less sensitive to the orientations of nuclei. At energies significantly higher than the Coulomb barrier ($E_{\rm c.m.}=860$~MeV or 10 MeV/u in our case) the orientational effects disappear.

The experiment on the $^{160}$Gd~+~$^{186}$W collisions was done at the CORSET setup in FLNR JINR~\cite{Kozulin17}. The angular ranges $25^{\circ}\leq\theta_{\rm lab.}\leq 65^{\circ}$ and $29^{\circ}\leq\theta_{\rm lab.}\leq 66^{\circ}$ were covered in the experiment to detect binary reaction products in coincidence respectively at $E_{\rm c.m.}=462$ and 502 MeV. Shaded areas in Fig.~\ref{img:GdWorientations} indicate the angular ranges set for detecting the TLFs. The TKEL, mass, and angular distributions of reaction products obtained in the experiment and the calculated ones are shown in Fig.~\ref{img:GdWresults}. The calculations were performed for the experimentally covered angular ranges. The experimental energy and mass resolutions (FWHM) of 12 MeV and 3 units were taken into account in the theoretical calculations. A satisfactory overall agreement with the data can be mentioned. Some overestimation of the experimental mass distribution is however seen for the collision energy $E_{\rm c.m.}=462$ MeV while a rather good description has been achieved for $E_{\rm c.m.}=502$ MeV.

It should be mentioned that the well-seen growth of the TLFs yields with increasing energy [panels (c) and (d) in Fig.~\ref{img:GdWresults}] is due to the strong dependence of the orientational effects on the collision energy, which is correctly reproduced by the present model.

\subsection{$^{208}$Pb~+~$^{208}$Pb/$^{238}$U systems}

\begin{figure}[t]
\centering
  \includegraphics[width=1\linewidth]{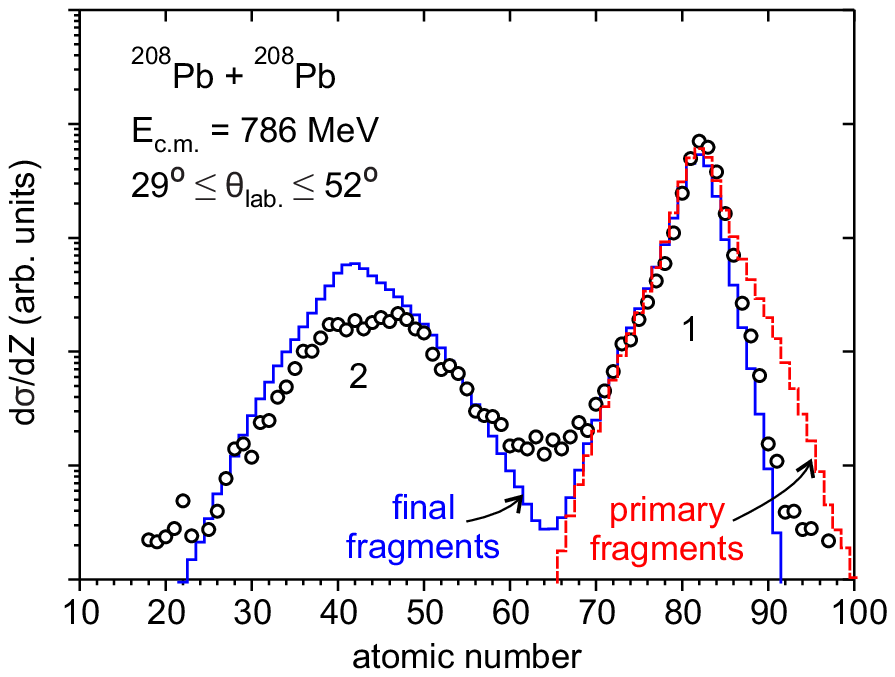}
  \caption{\small The charge distribution of final products obtained in the $^{208}$Pb~+~$^{208}$Pb reaction at energy $E_{\rm c.m.}=786$ MeV. The dashed and solid histograms represent calculated distributions of primary and final fragments, respectively. Symbols correspond to the experimental data~\cite{Tanabe80}. Contributions of DI (1) and SeqF (2) processes to inclusive charge distribution are shown.}
  \label{img:PbZdistr}
\end{figure}

\begin{figure*}[t]
\centering
  \includegraphics[width=0.8\linewidth]{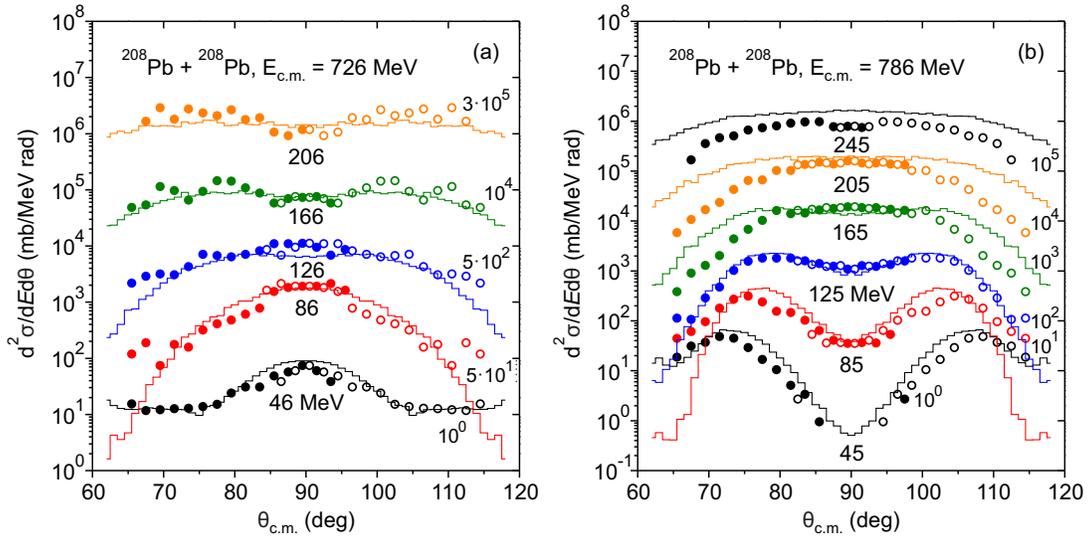}
  \caption{\small The angular distributions of primary products for different TKEL bins obtained in the $^{208}$Pb~+~$^{208}$Pb reaction at $E_{\rm c.m.}=726$ and 786 MeV. The histograms are calculation results and symbols correspond to the experimental data~\cite{Tanabe80}. The TKEL bins are 40 MeV wide, their midpoints and multiplication factors are indicated. }
  \label{img:PbAngles}
\end{figure*}

\begin{figure*}[!t]
\centering
  \includegraphics[width=1\textwidth]{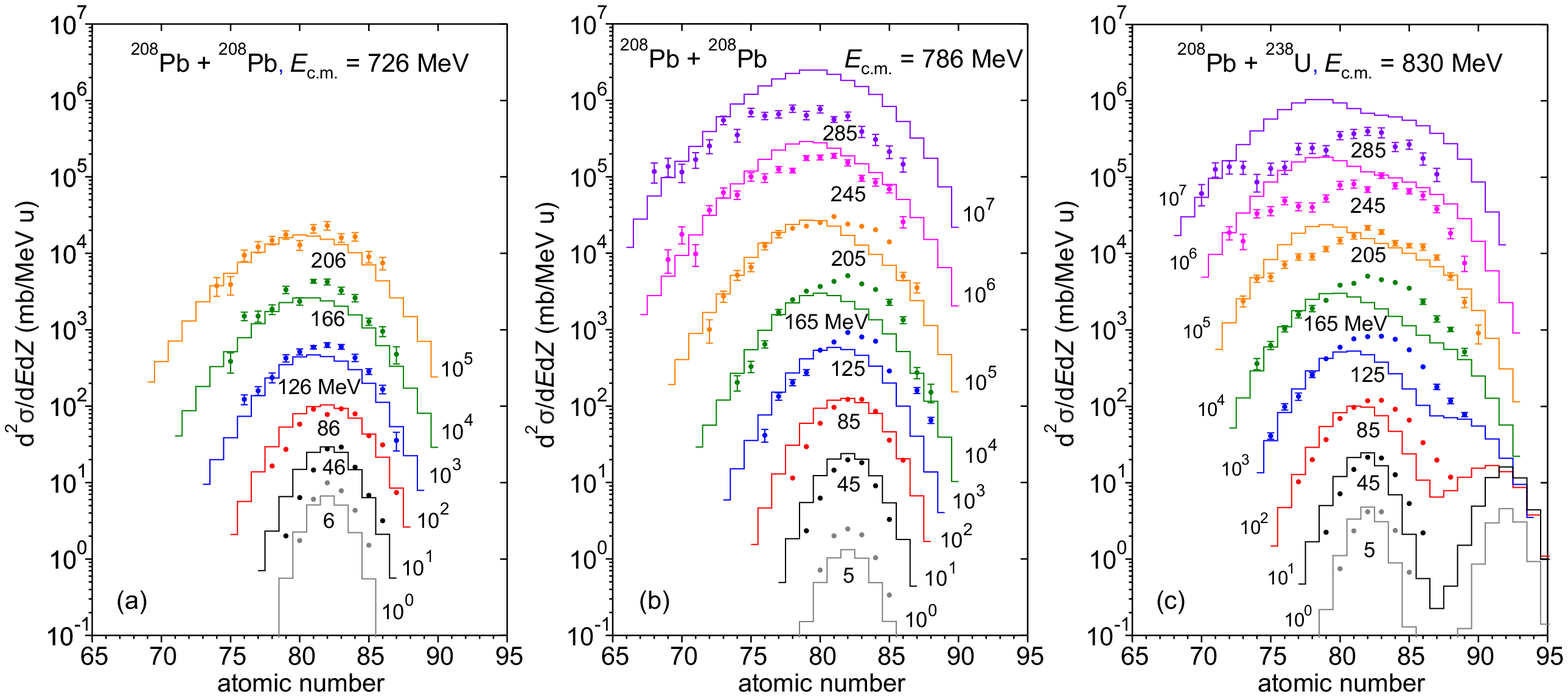}
  \caption{\small The charge distributions of final products for different TKEL values obtained in the $^{208}$Pb~+~$^{208}$Pb reaction at energies $E_{\rm c.m.}=726$ and 786 MeV (a,b) and the $^{208}$Pb~+~$^{238}$U reaction at energy $E_{\rm c.m.}=830$ MeV (c). Energy bins are 40 MeV wide, their midpoints and multiplication factors are indicated. The following angular ranges were considered both on the experimental data and on the calculations: $68^{\circ}\leq\theta_{\rm c.m.}\leq 90^{\circ}$ in the case of $^{208}$Pb~+~$^{208}$Pb and $70^{\circ}\leq\theta_{\rm c.m.}\leq 102^{\circ}$ in the case of $^{208}$Pb~+~$^{238}$U.
  The experimental data (symbols) are taken from Ref.~\cite{Tanabe80}.}
  \label{img:PbCharges}
\end{figure*}

In collisions of heavy nuclei the SeqF of highly excited heavy reaction products may play a significant role. It is not always possible to completely distinguish SeqF from the binary products, but it can be done quite reliably for the $^{208}$Pb~+~$^{208}$Pb reaction (see Fig.~\ref{img:PbZdistr}). The products with $Z>65$ form the main component of the charge distribution of quasi-elastic and DI collisions with the position of the maximum at $Z=82$. The additional component centered at $Z\approx40$ consists of SeqF fragments having $15\le Z\le 65$. The SeqF process hinders the survival of heavy above-target nuclei limiting their production in DI collisions.

The $^{208}$Pb~+~$^{208}$Pb reaction products were measured covering the $29^{\circ}\leq\theta_{\rm lab.}\leq 52^{\circ}$ angular range at two collision energies $E_{\rm c.m.}=726$ and 786 MeV~\cite{Tanabe80}. The experimental charge resolution of 2.5 units (FWHM) was taken into account in the calculations. The histograms were normalized to DI events and fit the experimental data rather well. The sequential fission rates were deduced experimentally as a ratio of the missing cross section (due to the SeqF process) to the one for primary products, subtracting the quasi-elastic events having TKEL$<$50 MeV. The found rates for the $^{208}$Pb~+~$^{208}$Pb reaction at collision energies $E_{\rm c.m.}=$726 and 786 MeV are respectively 11$\%$ and 16$\%$. The corresponding calculated values are 10$\%$ and 21$\%$. Obviously, the SeqF probability grows with increasing collision energy. Therefore, it seems reasonable to use the near-barrier collision energies in order to reduce the SeqF component and to increase the yields of above-target products.

The Coulomb force dominates over the attractive nuclear force in interaction of two heavy nuclei such as lead. It results in a broad angular distribution of the DI reaction products in contrast to the peaked distribution of fragments near the grazing angle obtained in the collisions of lighter nuclei such as Sm~+~Sm (Fig.~\ref{img:SmAngles}). The emission angles of the Pb~+~Pb reaction fragments continuously increase with increasing TKEL. This behavior is shown in Fig.~\ref{img:PbAngles} for the primary reaction products. The angular distributions of fragments for different TKEL values should be symmetric with respect to 90$^{\circ}$. The experimental data obtained in the $65^{\circ}\leq\theta_{\rm c.m.}\leq 95^{\circ}$ angular range for PLFs were doubled and symmetrically reflected to reproduce the missing component of TLFs.

\begin{figure*}[t]
\centering
  \includegraphics[width=0.9583\linewidth]{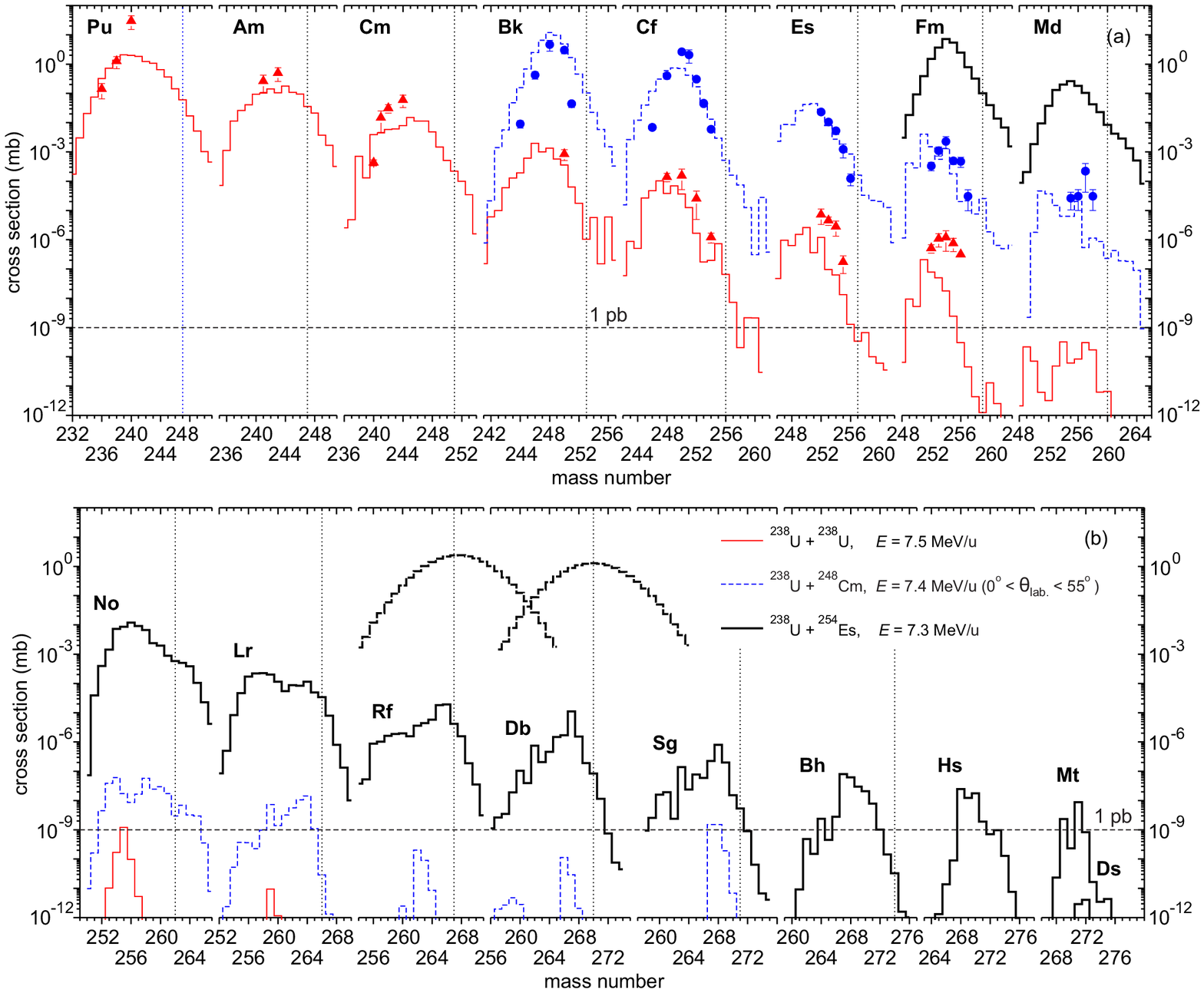}
  \caption{Isotopic distributions of final above-target products obtained in collisions of actinides. The thin, dashed, and thick histograms correspond to the results of the calculations for the reactions $^{238}$U~+~$^{238}$U ($E=7.5$ MeV/u), $^{238}$U~+~$^{248}$Cm ($E=7.4$ MeV/u), and $^{238}$U~+~$^{254}$Es ($E=7.3$ MeV/u), respectively. The experimental data for the~$^{238}$U~+~$^{238}$U reaction (triangles) are taken from Ref.~\cite{Schadel78}, and for $^{238}$U~+~$^{248}$Cm (circles) are from Ref.~\cite{Schadel82}. For more details see the text. The heaviest known isotopes of given chemical elements are indicated by the vertical dotted lines. The thick dashed curves show primary isotopic distributions of Rf and Db.}
\label{img:UUCmIsotopeDistr}
\end{figure*}

\begin{figure*}[t]
\centerline{%
\includegraphics[width=0.7\linewidth]{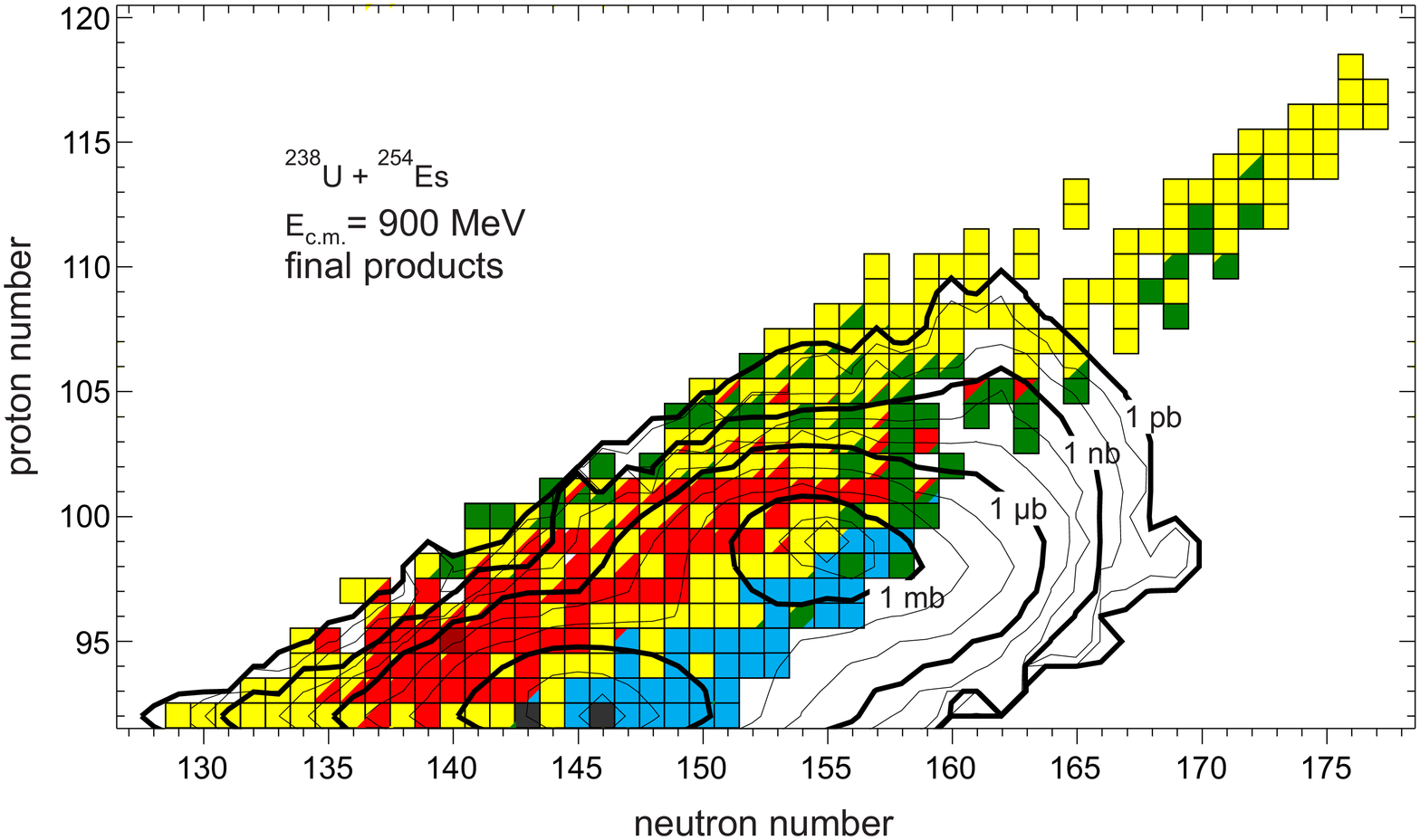}}
\caption{Production cross sections of final products in the $^{238}$U~+~$^{254}$Es reaction at $E_{\rm c.m.}=900$ MeV. Contour lines are drawn over an order of magnitude of the cross section down to 1 pb.}
\label{img:Map_UEs}
\end{figure*}

The charge distributions of fragments with low TKEL have maxima near the $\theta^{\rm gr}_{\rm c.m.}\approx90^{\circ}$ and 75$^{\circ}$ grazing angles for collision energies $E_{\rm c.m.}=726$ and 786 MeV, respectively. The part of the angular distribution corresponding to PLFs shifts to larger angles with increasing TKEL and crosses the symmetric branch for TLFs at $\theta_{\rm c.m.}=90^{\circ}$.

The $^{208}$Pb~+~$^{238}$U reaction was investigated under the same experimental conditions as for the $^{208}$Pb~+~$^{208}$Pb one~\cite{Tanabe80}. The charge distributions of products of the both reactions for different TKEL bins are compared in Fig.~\ref{img:PbCharges}. For this comparison the experimental data were integrated over the angular ranges
$68^{\circ}\leq\theta_{\rm c.m.}\leq 90^{\circ}$ in the case of the $^{208}$Pb~+~$^{208}$Pb reaction and
$70^{\circ}\leq\theta_{\rm c.m.}\leq 102^{\circ}$ in the case of the $^{208}$Pb~+~$^{238}$U reaction.

The calculations agree well with the experimental data on the panel (a) of Fig.~\ref{img:PbCharges} where the data for low collision energy are presented. However, it is seen that the calculated cross section maxima are slightly shifted towards the lower atomic numbers for larger TKEL values, while the maxima of the experimental distributions are almost unchanged. It becomes more evident at higher collision energies [panels (b) and (c)]. Since evaporation of protons from the primary fragments is suppressed, this effect is mainly due to the SeqF.

\subsection{$^{238}$U~+~$^{238}$U/$^{248}$Cm/$^{254}$Es systems}

Collisions of heavy actinides are of particular interest for the developed model. Actinides are deformed in their ground states and their orientations significantly influence the collision dynamics. The problem of synthesis of heavy and superheavy nuclei in DI collisions of actinides has been studied experimentally since the late 1970's for the $^{238}$U~+~$^{238}$U/$^{248}$Cm reactions~\cite{Schadel78, Schadel82, Kratz86}. The fact that final isotopic yields of heavy TLFs obtained in collisions involving $^{238}$U and $^{248}$Cm drop exponentially with increasing atomic number has been established in a series of these and other works (see, e.g.,~\cite{Hoffman85}). As a rule, the transfer of each proton towards the heavier reaction partner leads to a decrease of the corresponding production cross section by an order of magnitude. The results of our calculations shown in Fig.~\ref{img:UUCmIsotopeDistr} are in agreement with this trend.

Note, the experimental data shown in Fig.~\ref{img:UUCmIsotopeDistr} were obtained by radiochemical methods, and a thick target was used in the $^{238}$U~+~$^{238}$U experiment. Hence the final products of the $^{238}$U~+~$^{238}$U reaction shown by triangles in Fig.~\ref{img:UUCmIsotopeDistr} were detected in the whole angular range for collision energies $E\leq 7.5$ MeV/u. The calculations were performed for the fixed value of $E=7.5$ MeV/u). The $0^{\circ}\leq\theta_{\rm lab.}\leq 55^{\circ}$ angular range was covered in the experiment on the $^{238}$U~+~$^{248}$Cm collisions at $E=7.4$ MeV/u~\cite{Schadel78, Schadel82}.

A rather good agreement of the calculated and experimental isotopic yields was achieved for the absolute cross section values, and a small shift of the maxima of the calculated distributions towards lower masses can be seen in panel (a) of Fig.~\ref{img:UUCmIsotopeDistr} for the heaviest measured nuclides. Irregularities in the results of the calculations for low cross sections are caused by poor statistics. In order to improve agreement with the data shown in Fig.~\ref{img:UUCmIsotopeDistr}~(a) one should increase the nucleon transfer rates $\lambda_{A}^{0}$ and $\lambda_{Z}^{0}$ responsible in our model for the proton and neutron transfer probabilities in the vicinity of the contact point (see Eqs. (26) and (27) of Ref.~\cite{Karpov17}). The required increase should be about two times for the actinide-actinide collisions. Preliminary analysis performed for much lighter systems (e.g., $^{40}$Ca+$^{208}$Pb), which are not discussed here, requires, opposite, decreasing of the transfer rate coefficients to describe the existing data for nucleon transfer cross sections. Preliminarily, we found that the transfer rates should depend on the total mass of the system as $\sim A^2/1.2\,10^{5}$. We also found that such the change of the model parameters will not alter noticeably any result or conclusion made in this paper. Moreover, changes of the calculated quantities for lighter (than actinide-actinide) systems studied in this paper are even smaller and nearly invisible. Therefore, we decided not to alter the model parameters fixed in Ref.~\cite{Karpov17} and consider necessity of any change after detailed study of reactions between lighter ions in forthcoming works.

The production cross sections of nuclei with $Z>103$ do not exceed 1~pb in the $^{238}$U~+~$^{248}$Cm collisions, which makes the synthesis of the superheavy elements unreasonable using this reaction. However, in some cases the yields of yet-unknown neutron-enriched isotopes of heavy actinides are sufficiently large for their experimental identification. The vertical dotted lines in Fig.~\ref{img:UUCmIsotopeDistr} indicate the heaviest known isotopes for each chemical element.

The production cross sections of heavy TLFs for the given $Z$ are approximately four orders of magnitude larger for the $^{238}$U~+~$^{248}$Cm reaction compared with $^{238}$U~+~$^{238}$U. It certainly motivates one to use the heaviest available target in order to achieve the largest production cross sections for heavy and superheavy nuclei in DI collisions. We have performed calculations for the $^{238}$U~+~$^{254}$Es reaction as one of the possible combinations with heavy target~(see Figs.~\ref{img:UUCmIsotopeDistr} and \ref{img:Map_UEs}). The collision energy was set to $E=7.3$ MeV/u ($E_{\rm c.m.}=900$ MeV), which is slightly above the Coulomb barrier for the side-to-side mutual orientation ($V_C=874$ MeV). The whole angular range is covered in the calculations, but practically all above-target products are emitted in the forward angles up to $\theta_{\rm lab.}=55^{\circ}$.

The calculations for the $^{238}$U~+~$^{254}$Es reaction predict an additional shoulder for No and Lr isotopic distributions that becomes a maximum for Rf, Db and Sg distributions. The explanation of this phenomenon is the impact of the $N=162$ neutron sub-shell on formation of final reaction products from primary ones during the de-excitation process. Note, that there is no visible effect of this sub-shell on the formation of the primary products (compare primary and final fragments for Rf and Db isotopes in panel (b) of Fig.~\ref{img:UUCmIsotopeDistr}). Impact of the $N=162$ sub-shell is also present in Db and Sg distributions obtained in the $^{238}$U~+~$^{248}$Cm reaction.

As can be seen from Fig.~\ref{img:UUCmIsotopeDistr}, the yields of primary fragments are rather large. For example, excited $Z=114$ nuclei can be produced with the cross sections of about 1~${\rm \mu}$b (not shown in Fig.~\ref{img:UUCmIsotopeDistr}). Nevertheless, high excitation energies and angular momenta lead to rather low probabilities of their survival.

The production cross sections for the $^{238}$U~+~$^{254}$Es reaction products with $Z > 91$ are shown in Fig.~\ref{img:Map_UEs}. In this reaction unknown neutron-enriched isotopes of elements from U to Md can be produced with the cross sections exceeding 1~${\rm \mu}$b. The above-discussed decrease of the isotopic distributions with increasing atomic number imposes certain restrictions on the formation of above-target nuclei. In particular, the possibility of synthesis of unknown superheavy nuclides in DI collisions of actinides is rather limited.

The initial orientation of statically deformed nuclei also affects the production yields of heavy above-target nuclei. Lower excitation energies of primary fragments formed in a more compact side-to-side collisions will increase their survival probability. On the other hand, the cross sections for primary products for the side-to-side collisions are smaller than for other orientations. The final yield is a product of the survival probability and primary cross section. Determination of an optimal collision energy is of great importance for planning experiments on production of heavy nuclei and will be a topic of future studies.

\section{Conclusions and outlook}

In this paper, the multinucleon transfer processes in low-energy collisions are analyzed for both spherical and statically deformed nuclei. The model provided a reasonable agreement between the calculated and the measured energy, angular, charge, and isotopic distributions of reaction products for a number of MNT reactions with medium-mass and heavy nuclei.

The mutual orientation of colliding statically deformed nuclei in the entrance channel strongly affects the reaction dynamics at near-barrier energies. This applies to the absolute values and widths of the energy, angular, mass, and charge distributions of reaction products obtained for different mutual orientations of projectile and target nuclei. These orientational effects gradually disappear with increasing collision energy to the values well above the Coulomb barrier for all orientations.

The developed approach allows us to calculate yields of the above-target nuclei produced in collisions of heavy actinides at near-barrier collision energies. The calculation results show a strong exponential drop of the production cross sections with increasing atomic number due to high excitation energies and angular momenta of primary products. This drop was earlier observed experimentally for the $^{238}$U~+~$^{238}$U/$^{248}$Cm reactions \cite{Schadel78,Schadel82}. This fact makes the region of new superheavy nuclides to be hardly reachable in MNT reactions. However, there is a real chance to produce a number of neutron-enriched isotopes of heavy actinides with the cross sections exceeding 1~${\rm \mu}$b in the MNT reaction with the $^{254}$Es target.

Both theoretical and experimental studies of the energy dependence of the production yields of heavy neutron-enriched nuclei in MNT reactions with heavy ions is of special interest for determining conditions for their synthesis.

\begin{acknowledgments}
One of us (V.V.S.) was supported by JINR grant for young scientists and specialists No. 18-502-08.
\end{acknowledgments}


\begin{thebibliography}{10}

\bibitem{Artukh73}
A.~G. Artukh, G.~F. Gridnev, V.~L. Mikheev, V.~V. Volkov, and J.~Wilczynski.
\newblock Transfer reactions in the interaction of
  ${}^{40}\mathrm{Ar}+{}^{232}\mathrm{Th}$.
\newblock {\em Nuclear Physics A}, 215(1):91 -- 108, 1973.

\bibitem{Dasso94}
C.~H. Dasso, G.~Pollarolo, and A.~Winther.
\newblock Systematics of isotope production with radioactive beams.
\newblock {\em Phys. Rev. Lett.}, 73:1907--1910, Oct 1994.

\bibitem{Broda06}
R.~Broda.
\newblock Spectroscopic studies with the use of deep-inelastic heavy-ion
  reactions.
\newblock {\em Journal of Physics G: Nuclear and Particle Physics}, 32(6):R151,
  2006.

\bibitem{ZagrebaevGreiner07_GdW}
V.~Zagrebaev and W.~Greiner.
\newblock Shell effects in damped collisions: a new way to superheavies.
\newblock {\em Journal of Physics G: Nuclear and Particle Physics},
  34(11):2265, 2007.

\bibitem{ZagrebaevGreiner08}
V.~Zagrebaev and W.~Greiner.
\newblock New way for the production of heavy neutron-rich nuclei.
\newblock {\em Journal of Physics G: Nuclear and Particle Physics},
  35(12):125103, 2008.

\bibitem{Corradi09}
L.~Corradi, G.~Pollarolo, and S.~Szilner.
\newblock Multinucleon transfer processes in heavy-ion reactions.
\newblock {\em Journal of Physics G: Nuclear and Particle Physics},
  36(11):113101, 2009.

\bibitem{Karpov17}
A.~V. Karpov and V.~V. Saiko.
\newblock Modeling near-barrier collisions of heavy ions based on a
  langevin-type approach.
\newblock {\em Phys. Rev. C}, 96:024618, Aug 2017.

\bibitem{Kozulin12}
E.~M. Kozulin, E.~Vardaci, G.~N. Knyazheva, A.~A. Bogachev, S.~N. Dmitriev,
  I.~M. Itkis, M.~G. Itkis, A.~G. Knyazev, T.~A. Loktev, K.~V. Novikov, E.~A.
  Razinkov, O.~V. Rudakov, S.~V. Smirnov, W.~Trzaska, and V.~I. Zagrebaev.
\newblock Mass distributions of the system
  ${}^{136}\mathrm{Xe}+{}^{208}\mathrm{Pb}$ at laboratory energies around the
  coulomb barrier: A candidate reaction for the production of neutron-rich
  nuclei at ${N}=126$.
\newblock {\em Phys. Rev. C}, 86:044611, Oct 2012.

\bibitem{Barrett15}
J.~S. Barrett, W.~Loveland, R.~Yanez, S.~Zhu, A.~D. Ayangeakaa, M.~P.
  Carpenter, J.~P. Greene, R.~V.~F. Janssens, T.~Lauritsen, E.~A. McCutchan,
  A.~A. Sonzogni, C.~J. Chiara, J.~L. Harker, and W.~B. Walters.
\newblock $^{136}\mathrm{Xe}+^{208}\mathrm{Pb}$ reaction: A test of models of
  multinucleon transfer reactions.
\newblock {\em Phys. Rev. C}, 91:064615, Jun 2015.

\bibitem{Watanabe15}
Y.~X. Watanabe, Y.~H. Kim, S.~C. Jeong, Y.~Hirayama, N.~Imai, H.~Ishiyama,
  H.~S. Jung, H.~Miyatake, S.~Choi, J.~S. Song, E.~Clement, G.~de~France,
  A.~Navin, M.~Rejmund, C.~Schmitt, G.~Pollarolo, L.~Corradi, E.~Fioretto,
  D.~Montanari, M.~Niikura, D.~Suzuki, H.~Nishibata, and J.~Takatsu.
\newblock Pathway for the production of neutron-rich isotopes around the
  ${N}=126$ shell closure.
\newblock {\em Phys. Rev. Lett.}, 115:172503, Oct 2015.

\bibitem{Kozulin17}
E.~M. Kozulin, V.~I. Zagrebaev, G.~N. Knyazheva, I.~M. Itkis, K.~V. Novikov,
  M.~G. Itkis, S.~N. Dmitriev, I.~M. Harca, A.~E. Bondarchenko, A.~V. Karpov,
  V.~V. Saiko, and E.~Vardaci.
\newblock Inverse quasifission in the reactions
  $^{156,160}\mathrm{Gd}+^{186}\mathrm{W}$.
\newblock {\em Phys. Rev. C}, 96:064621, Dec 2017.

\bibitem{Iwamoto96}
A.~Iwamoto, P.~M{\"o}ller, J.~R. Nix, and H.~Sagawa.
\newblock Collisions of deformed nuclei: A path to the far side of the
  superheavy island.
\newblock {\em Nuclear Physics A}, 596(2):329 -- 354, 1996.

\bibitem{Hinde96}
D.~J. Hinde, M.~Dasgupta, J.~R. Leigh, J.~C. Mein, C.~R. Morton, J.~O. Newton,
  and H.~Timmers.
\newblock Conclusive evidence for the influence of nuclear orientation on
  quasifission.
\newblock {\em Phys. Rev. C}, 53:1290--1300, Mar 1996.

\bibitem{Nishio00}
K.~Nishio, H.~Ikezoe, S.~Mitsuoka, and J.~Lu.
\newblock Fusion of deformed nuclei in the reactions of
  ${}^{76}\mathrm{Ge}{+}^{150}\mathrm{Nd}$ and
  ${}^{28}\mathrm{Si}{+}^{198}\mathrm{Pt}$ at the coulomb barrier region.
\newblock {\em Phys. Rev. C}, 62:014602, May 2000.

\bibitem{Mitsuoka00}
S.~Mitsuoka, H.~Ikezoe, K.~Nishio, and J.~Lu.
\newblock Sub-barrier fusion of deformed nuclei in
  ${}^{60}\mathrm{Ni}{+}^{154}\mathrm{Sm}$ and
  ${}^{32}\mathrm{S}{+}^{182}\mathrm{W}$ reactions.
\newblock {\em Phys. Rev. C}, 62:054603, Oct 2000.

\bibitem{SaikoKarpov18}
V.~Saiko and A.~Karpov.
\newblock Dynamics of near-barrier collisions of statically deformed nuclei.
\newblock {\em Acta Physica Polonica B}, 49:307--312, 2018.

\bibitem{Schadel78}
M.~Sch\"adel, J.~V. Kratz, H.~Ahrens, W.~Br\"uchle, G.~Franz, H.~G\"aggeler,
  I.~Warnecke, G.~Wirth, G.~Herrmann, N.~Trautmann, and M.~Weis.
\newblock Isotope distributions in the reaction of $^{238}\mathrm{U}$ with
  $^{238}\mathrm{U}$.
\newblock {\em Phys. Rev. Lett.}, 41:469--472, Aug 1978.

\bibitem{Schadel82}
M.~Sch\"adel, W.~Br\"uchle, H.~G\"aggeler, J.~V. Kratz, K.~S\"ummerer,
  G.~Wirth, G.~Herrmann, R.~Stakemann, G.~Tittel, N.~Trautmann, J.~M. Nitschke,
  E.~K. Hulet, R.~W. Lougheed, R.~L. Hahn, and R.~L. Ferguson.
\newblock Actinide production in collisions of $^{238}\mathrm{U}$ with
  $^{248}\mathrm{Cm}$.
\newblock {\em Phys. Rev. Lett.}, 48:852--855, Mar 1982.

\bibitem{Golabek10}
C.~Golabek, S.~Heinz, W.~Mittig, F.~Rejmund, A.~C.~C. Villari,
  S.~Bhattacharyva, D.~Boilley, G.~De~France, A.~Drouart, L.~Gaudefroy,
  L.~Giot, V.~Maslov, M.~Morjean, G.~Mukherjee, Yu. Penionzkevich,
  P.~Roussel-Chomaz, and C.~Stodel.
\newblock Investigation of deep inelastic reactions in $^{238}\mathrm{U} +
  ^{238}\mathrm{U}$ at {Coulomb} barrier energies.
\newblock {\em The European Physical Journal A}, 43(3):251--259, Mar 2010.

\bibitem{Zagrebaev11}
V.~I. Zagrebaev, A.~V. Karpov, I.~N. Mishustin, and Walter Greiner.
\newblock Production of heavy and superheavy neutron-rich nuclei in neutron
  capture processes.
\newblock {\em Phys. Rev. C}, 84:044617, Oct 2011.

\bibitem{Zhao13}
Kai Zhao, Zhuxia Li, Xizhen Wu, and Yingxun Zhang.
\newblock Production probability of superheavy fragments at various initial
  deformations and orientations in the $^{238}\mathrm{U}{+}^{238}\mathrm{U}$
  reaction.
\newblock {\em Phys. Rev. C}, 88:044605, Oct 2013.

\bibitem{Zhao16}
Kai Zhao, Zhuxia Li, Yingxun Zhang, Ning Wang, Qingfeng Li, Caiwan Shen,
  Yongjia Wang, and Xizhen Wu.
\newblock Production of unknown neutron-rich isotopes in
  $^{238}\text{U}+\phantom{\rule{0.16em}{0ex}}^{238}\text{U}$ collisions at
  near-barrier energy.
\newblock {\em Phys. Rev. C}, 94:024601, Aug 2016.

\bibitem{Feng_UU09}
Zhao-Qing Feng, Gen-Ming Jin, and Jun-Qing Li.
\newblock Production of heavy isotopes in transfer reactions by collisions of
  $^{238}\mathrm{U}+^{238}\mathrm{U}$.
\newblock {\em Phys. Rev. C}, 80:067601, Dec 2009.

\bibitem{Oberacker14}
V.~E. Oberacker, A.~S. Umar, and C.~Simenel.
\newblock Dissipative dynamics in quasifission.
\newblock {\em Phys. Rev. C}, 90:054605, Nov 2014.

\bibitem{Umar16}
A.~S. Umar, V.~E. Oberacker, and C.~Simenel.
\newblock Fusion and quasifission dynamics in the reactions
  $^{48}\mathrm{Ca}+^{249}\mathrm{Bk}$ and $^{50}\mathrm{Ti}+^{249}\mathrm{Bk}$
  using a time-dependent hartree-fock approach.
\newblock {\em Phys. Rev. C}, 94:024605, Aug 2016.

\bibitem{Maruhn72}
J.~Maruhn and W.~Greiner.
\newblock The asymmetric two center shell model.
\newblock {\em Zeitschrift f{\"u}r Physik}, 251(5):431--457, Oct 1972.

\bibitem{Migdal83}
A.B. Migdal.
\newblock {\em Theory of finite Fermi systems and applications to atomic
  nuclei}.
\newblock NY : Interscience, 1967.

\bibitem{ZagrebaevKarpov07}
V.~Zagrebaev, A.~Karpov, Y.~Aritomo, M.~Naumenko, and W.~Greiner.
\newblock Potential energy of a heavy nuclear system in fusion-fission
  processes.
\newblock {\em Physics of Particles and Nuclei}, 38(4):469--491, Jul 2007.

\bibitem{NRV}
A.~V. Karpov A. P. Alekseev M. A. Naumenko V. A. Rachkov V. V.~Samarin
  V.~I.~Zagrebaev, A. S.~Denikin and V.~V. Saiko.
\newblock {NRV} web knowledge base on low-energy nuclear physics,
  http://nrv.jinr.ru.

\bibitem{Moller16}
P.~M{\"o}ller, A.J. Sierk, T.~Ichikawa, and H.~Sagawa.
\newblock Nuclear ground-state masses and deformations: Frdm(2012).
\newblock {\em Atomic Data and Nuclear Data Tables}, 109–110:1--204, 2016.

\bibitem{Strutinsky67}
V.M. Strutinsky.
\newblock Shell effects in nuclear masses and deformation energies.
\newblock {\em Nuclear Physics A}, 95(2):420 -- 442, 1967.

\bibitem{Strutinsky68}
V.M. Strutinsky.
\newblock Shells in deformed nuclei.
\newblock {\em Nuclear Physics A}, 122(1):1 -- 33, 1968.

\bibitem{Davies76}
K.~T.~R. Davies, A.~J. Sierk, and J.~R. Nix.
\newblock Effect of viscosity on the dynamics of fission.
\newblock {\em Phys. Rev. C}, 13:2385--2403, Jun 1976.

\bibitem{Sierk80}
A.~J. Sierk and J.~R. Nix.
\newblock Fission in a wall-and-window one-body-dissipation model.
\newblock {\em Phys. Rev. C}, 21:982--987, Mar 1980.

\bibitem{KarpovEPJA02}
{A. V. Karpov} and {G. D. Adeev}.
\newblock Langevin description of charge fluctuations in fission of highly
  excited nuclei.
\newblock {\em Eur. Phys. J. A}, 14(2):169--178, 2002.

\bibitem{KarpovNRV16}
A.~V. Karpov, A.~S. Denikin, A.~P. Alekseev, V.~I. Zagrebaev, V.~A. Rachkov,
  M.~A. Naumenko, and V.~V. Saiko.
\newblock {NRV} web knowledge base on low-energy nuclear physics.
\newblock {\em Physics of Atomic Nuclei}, 79(5):749--761, Sep 2016.

\bibitem{KarpovNRV17}
A.V. Karpov, A.S. Denikin, M.A. Naumenko, A.P. Alekseev, V.A. Rachkov, V.V.
  Samarin, V.V. Saiko, and V.I. Zagrebaev.
\newblock {NRV} web knowledge base on low-energy nuclear physics.
\newblock {\em Nuclear Instruments and Methods in Physics Research Section A:
  Accelerators, Spectrometers, Detectors and Associated Equipment},
  859(Supplement C):112 -- 124, 2017.

\bibitem{GEF}
K.-H. Schmidt, B.~Jurado, C.~Amouroux, and C.~Schmitt.
\newblock General description of fission observables: {GEF} model code.
\newblock {\em Nuclear Data Sheets}, 131(Supplement C):107 -- 221, 2016.
\newblock Special Issue on Nuclear Reaction Data.

\bibitem{WuHildenbrand81}
E.~C. Wu, K.~D. Hildenbrand, H.~Freiesleben, A.~Gobbi, A.~Olmi, H.~Sann, and
  U.~Lynen.
\newblock Influence of shell structure on neutron and proton exchange in the
  reactions of $^{144}\mathrm{Sm}$ on $^{144}\mathrm{Sm}$ and
  $^{154}\mathrm{Sm}$ on $^{154}\mathrm{Sm}$.
\newblock {\em Phys. Rev. Lett.}, 47:1874--1877, Dec 1981.

\bibitem{Hildenbrand83}
K.~D. Hildenbrand, H.~Freiesleben, A.~Gobbi, U.~Lynen, A.~Olmi, H.~Sann, and
  E.~C. Wu.
\newblock On the influence of shell structure in dissipative collisions.
\newblock {\em Nuclear Physics A}, 405(1):179 -- 204, 1983.

\bibitem{Tanabe80}
T.~Tanabe, R.~Bock, M.~Dakowski, A.~Gobbi, H.~Sann, H.~Stelzer, U.~Lynen,
  A.~Olmi, and D.~Pelte.
\newblock The {Pb-Pb} collision.
\newblock {\em Nuclear Physics A}, 342(1):194 -- 212, 1980.

\bibitem{Kratz86}
J.~V. Kratz, W.~Br\"uchle, H.~Folger, H.~G\"aggeler, M.~Sch\"adel,
  K.~S\"ummerer, G.~Wirth, N.~Greulich, G.~Herrmann, U.~Hickmann, P.~Peuser,
  N.~Trautmann, E.~K. Hulet, R.~W. Lougheed, J.~M. Nitschke, R.~L. Ferguson,
  and R.~L. Hahn.
\newblock Search for superheavy elements in damped collisions between
  $^{238}\mathrm{U}$ and $^{248}\mathrm{Cm}$.
\newblock {\em Phys. Rev. C}, 33:504--508, Feb 1986.

\bibitem{Hoffman85}
D.~C. Hoffman, M.~M. Fowler, W.~R. Daniels, H.~R. von Gunten, D.~Lee, K.~J.
  Moody, K.~Gregorich, R.~Welch, G.~T. Seaborg, W.~Br\"uchle, M.~Br\"ugger,
  H.~Gaggeler, M.~Schadel, K.~S\"ummerer, G.~Wirth, Th. Blaich, G.~Herrmann,
  N.~Hildebrand, J.~V. Kratz, M.~Lerch, and N.~Trautmann.
\newblock Excitation functions for production of heavy actinides from
  interactions of $^{40}\mathrm{Ca}$ and $^{48}\mathrm{Ca}$ ions with
  $^{248}\mathrm{Cm}$.
\newblock {\em Phys. Rev. C}, 31:1763--1769, May 1985.

\end{thebibliography}
\end{document}